\documentclass[11pt]{iopart}
\usepackage{bm}
\usepackage{latexsym}
\usepackage{dcolumn}
\usepackage{graphicx}
\usepackage{epsfig}
\usepackage{psfrag}
\usepackage{iopams}

\def\be {\begin{equation}}
\def\ee {\end{equation}}
\def\bea {\begin{eqnarray}}
\def\eea {\end{eqnarray}}
\def\bc {\begin{center}}
\def\ec {\end{center}}
\def\bfg {\begin{figure}}
\def\efg {\end{figure}}
\def\bi {\begin{itemize}}
\def\ei {\end{itemize}}
\def\nn {\nonumber}

\def\la {\label}
\def\le {\left}
\def\ri {\right}
\def\pa {\partial}
\def\fr {\frac}
\def\sq {\sqrt}

\def\a  {\alpha}
\def\b  {\beta}

\def\d  {\delta}
\def\D  {\Delta}

\def\f {\phi}
\def\F  {\Phi}
\def\k  {\kappa}
\def\l  {\lambda}
\def\L  {\Lambda}
\def\m  {\mu}
\def\n  {\nu}

\def\O  {\Omega}
\def\p  {\pi}
\def\r  {\rho}

\def\s {\sigma}

\def\vph {\varphi}
\def\vth {\vartheta}

\def\rx {\rho_{_X}}
\def\drx {\dot{\rho}_{_X}}
\def\px {p_{_X}}
\def\wx {w_{_X}}
\def\cx {c_{_X}}

\def\wI {w_{_I}}
\def\rI {\rho_{_I}}
\def\rJ {\rho_{_J}}

\def\SIJ {S_{_{IJ}}}
\def\sigI {\sigma_{_I}}
\def\sigJ {\sigma_{_J}}
\def\sigIJ {\sigma_{_{IJ}}}
\def\drI {\dot{\rho}_{_I}}
\def\pI {P_{_I}}
\def\pJ {P_{_J}}
\def\dpI {\dot{P}_{_I}}
\def\fI {\phi_{_I}}
\def\fJ {\phi_{_J}}
\def\dfI {\dot{\phi}_{_I}}
\def\xI {X_{_I}}
\def\yI {y_{_I}}
\def\vI {v_{_I}}
\def\vJ {v_{_J}}
\def\vIJ {v_{_{IJ}}}
\def\eI {e_{_I}}
\def\eJ {e_{_J}}
\def\eIJ {e_{_{IJ}}}
\def\piI {\pi_{_I}}
\def\ya {y_{_0}}
\def\yb {y_{_1}}
\def\yc {y_{_2}}
\def\cI {c_{_I}}
\def\cJ {c_{_J}}
\def\cIJ {c_{_{IJ}}}
\def\etaI {\eta_{_I}}
\def\ctI {\tilde{c}_{_I}}

\def\cbIJ {{\bar c}_{_{IJ}}}

\def\nuI {\nu_{_I}}
\def\aa {\alpha_{_0}}
\def\ab {\alpha_{_1}}
\def\ac {\alpha_{_2}}
\def\Aa {A_{_0}}
\def\Ab {A_{_1}}
\def\Ac {A_{_2}}
\def\An {A_{_{N-1}}}
\def\tt {\tilde{t}}

\begin{document}

\title{Multiple kinetic k-essence, phantom barrier crossing and stability}

\author{Sourav Sur and Saurya Das}

\address{Dept. of Physics, University of Lethbridge \\
4401 University Drive, Lethbridge, Alberta, Canada T1K 3M4}

\eads{sourav.sur@uleth.ca and saurya.das@uleth.ca}

\begin{abstract}
We investigate models of dark energy with purely kinetic multiple k-essence
sources that allow for the crossing of the phantom divide line, without violating
the conditions of stability. It is known that with more than one kinetic k-field
one can possibly construct dark energy models whose equation of state parameter
$\wx$ crosses $-1$ (the phantom barrier) at recent red-shifts, as indicated by
the Supernova Ia and other observational probes. However, such models may suffer
from cosmological instabilities, as the effective speed of propagation $\cx$ of
the dark energy density perturbations may become {\it imaginary} while the $\wx =
-1$ barrier is crossed. Working out the expression for $\cx$ we show that multiple
kinetic k-essence fields do indeed lead to a $\wx = -1$ crossing dark energy model,
satisfying the stability criterion $\cx^2 \geq 0$ as well as the condition $\cx
\leq 1$ (in natural units), which implies that the dark energy is not super-luminal.
As a specific example, we construct a phantom barrier crossing model involving three
k-fields for which $\cx$ is a constant, lying between $0$ and $1$. The model fits well
with the latest Supernova Ia Union data, and the best fit shows that $\wx$ crosses
$-1$ at red-shift $z \sim 0.2$, whereas the dark energy density nearly tracks the
matter density at higher red-shifts.
\end{abstract}




\maketitle


\section{Introduction       \la{sec:intro}}

The late-time accelerated expansion of the universe is now well established through
the observations of the luminosity distance -- red-shift relation for type Ia Supernovae
(SNIa) \cite{perl,tonry,riess1,riess2,astier,davis,kowal,rubin}. Measurements of the cosmic
microwave background (CMB) temperature fluctuations by the Wilkinson Microwave Anisotropy
Probe (WMAP) \cite{wmap1,wmap2} as well as the large scale red-shift data from the Sloan
Digital Sky Survey (SDSS) \cite{sdss} further indicate that the universe is very nearly
spatially flat and consists of about $70 \%$ of dark energy (DE) which drives the cosmic
acceleration. The value of the DE equation of state (EoS) parameter, i.e., the ratio of
pressure to energy density of DE, $\wx = \px/\rx$, should be less than $-1/3$ to maintain
this acceleration. Although the simplest candidate for DE is a cosmological constant ($\L$),
for which $\wx = -1$, there are serious theoretical problems associated with it \cite{lambda}.
$\L$ needs to be extremely {\it fine tuned} so that the DE at present is small compared to
the Planck scale. A fine tuning is also required to keep the ($\L$-) dark energy density
$\rx$ comparable with the present critical density $\r_{0c}$, while the acceleration begins
only in the recent past --- the {\it coincidence problem}. From the observational point of
view, although there are indications that $\wx$ is very close to $-1$ at present, combined
studies of SNIa, CMB and baryon acoustic oscillation (BAO) peaks do actually point toward
a {\it dynamical} DE with EoS parameter $\wx$ slowly varying with red-shift $z$. More
specifically, the so-called $\L$CDM model, which consists of $\L$ and cold dark matter as
dominant components, although consistent with the low-$z$ observational data, does not
produce sufficiently good fits with the data at relatively higher $z$ ($\gtrsim 1$).
Accordingly, there have been a large number of model-independent and model-dependent DE
parameterizations explored in the literature \cite{cpl,alam,visser,gong,param}. Extensive
studies reveal that in parameterizing the DE to produce good fits with the observational
data, it is always desirable to have no inherent restriction imposed on $\wx$ from the
underlying theory. For instance, it is not desirable to have $\wx$ always $\geq -1$, which
is required for the theoretical consistency of many popular DE models, such as quintessence,
tachyon, dilaton, etc. For the reviews on various dynamical aspects of DE, see \cite{de-rev}.

Within the scope of minimally coupled canonical scalar field DE models, such as quintessence
\cite{quin}, the crossing from $\wx > -1$ to $\wx < -1$ can never happen and there is a need
of more than one scalar field, at least one of which is of {\it phantom} nature, i.e., carries
a wrong sign in front of the kinetic term \cite{phantom}. For this reason the $\wx = -1$ line
is commonly known as the phantom divide line (PDL). Phantom fields violate the dominant energy
condition, and therefore could give rise to classical instabilities \cite{cht}. Moreover,
since the phantom energy density is unbounded from below, the quantum mechanical stability
is in jeopardy. The vacuum is unstable against the decay into ghosts and positive energy
particles, which couple to gravitons producing further decay into cosmic gamma rays \cite{cline}.
In an expanding Friedmann universe, the extremely repulsive (anti-gravitating) nature of the
phantom fields pose more problems as the phantom energy density generically tends to infinity in
finite time in future leading to the so-called ``Big Rip" singularity \cite{cald}, as well as
some other mild singularities. For a general categorization of these singularities see for
instance \cite{noj}.

Apart from the coupled quintessence and phantom models, commonly known as the {\it quintom}
models \cite{quintom}, there are various other models in which the phantom barrier crossing is
shown to be achieved in a consistent way. Notable among these are the scalar-tensor models
\cite{polarski}, brane-world models \cite{sahni}, string-inspired dilatonic ghost condensate
models \cite{dil}, quantum-corrected Klein-Gordon models with quartic potential \cite{onemli},
coupled DE models \cite{coupDE}, modified gravity models \cite{modgrav}, H-essence (complex
scalar) models \cite{wei}, etc. In this work, we focus on the so-called k-essence models
\cite{k-ess1,k-ess2}, which incorporate a natural generalization of quintessence via a Lagrangian
with non-linear dependence on the kinetic term(s) for the scalar field(s). For such models $\wx$
is not constrained to be $\geq -1$ and may in principle cross the PDL. However, it has been
argued by Vikman \cite{vik} that in the case of a single field k-essence model, with a Lagrangian
of the form $P(\f,X)$, where $X = -\fr 1 2 [\nabla \f]^2$ is the kinetic term of the scalar field
$\f$, the PDL crossing either leads to instabilities against cosmological perturbations or is
realized by a discrete set of phase space trajectories\footnote{There are however some
counter-arguments and counter-examples set in \cite{odint,andrianov} and it is shown that for a
generalized single field k-essence Lagrangian of the form $P(\f,X) = f(\f) X - V(\f)$ stable PDL
crossing is indeed possible for some specific functional form of $f(\f)$ and $V(\f)$.}. For a
multiple k-essence model, involving more than one k-fields $\fI ~(I=0,\dots,N-1)$, with kinetic
terms $\xI = -\fr 1 2 [\nabla \fI]^2$, the PDL crossing is shown to be possible even when the
Lagrangian is purely a function of the kinetic terms, $P = \sum \pI(\xI)$ \cite{chim}. However,
the stability of such a model against cosmological perturbations have not been explored in the
literature.

A major difficulty in dealing with the equations of scalar-type cosmological perturbations for
multiple scalar fields is that these equations are extremely coupled in the curvature and
isocurvature modes, even when the scalar fields are only minimally coupled to gravity
\cite{hwang,hwangnoh1,kodamasasaki,maartens}. For multiple k-essence, the perturbation equations
are complicated further, however there are some specific cases in which the expression for the
speed of propagation $\cx$ of the curvature perturbations could be obtained and the stability
analysis could be performed \cite{langlois,arroja}. For example, when the k-essence Lagrangian
is $P({\mathbf \F},X)$, where ${\mathbf \F} = (\f_{_0}, \f_{_1}, \cdots)$ and $X = - \fr 1 2
{\cal G}^{IJ} ({\mathbf \F}) \nabla_\m \fI \nabla^\m \fJ$ \cite{langlois}, the propagation speed
is simply given by the well-known result of Garriga and Mukhanov \cite{gm}, i.e., $\cx^2 =
\fr{\pa P}{\pa X} {\big /} \fr{\pa \rx}{\pa X}$. On the other hand, in the multi-field
Dirac-Born-Infeld (DBI) inflationary scenario, the propagation speed for the Lagrangian $P = -
f^{-1} ({\mathbf \F}) \sqrt{1 - 2 X f ({\mathbf \F})} - V ({\mathbf \F})$, is given in a
homogeneous background by $\cx = \sqrt{1 - 2 X f ({\mathbf \F})}$ \cite{arroja}. However, none
of these expressions for $\cx$ will hold for the most general Lagrangian $P = \sum \pI(\fI,\xI)$,
corresponding to the $N$ non-interacting fields ${\mathbf \F} = (\f_{_0}, \f_{_1}, \cdots)$. A
comprehensive study of the curvature perturbation equations reveals that one can, in principle,
obtain the expression of $\cx$ for the multi k-field Lagrangian $P = \sum \pI(\fI,\xI)$ \cite{ss},
leaving aside the isocurvature modes which couple weakly with the curvature mode in the large-scale
limit \cite{hwangnoh1,kodamasasaki}.

In this paper we first investigate whether the multiple kinetic k-essence DE models that allow
the PDL crossing is cosmologically {\it stable}, i.e., the speed of propagation $\cx$ of the DE
density perturbations satisfies the condition $\cx^2 \geq 0$. It is also relevant to check
whether $\cx \leq 1$ (in natural units), so that the DE perturbations do not travel faster than
light \cite{bonvin}. However, the requirement $\cx \leq 1$ is not potentially stringent, because
it has been argued that causal paradoxes do not necessarily arise even when $\cx > 1$ \cite{brun}.
We find that in most circumstances (including the model described in \cite{chim}) where $\cx$
dynamically varies with time, the PDL crossing is associated with a flip of sign of $\cx^2$, thus
giving rise to violent instability. An appropriate choice for $\cx^2$ in a stable DE model is
therefore a {\it constant} whose value lies between $0$ and $1$. Imposing this condition on $\cx^2$
we work out the constraint on the field configuration. As a specific example we try to construct a
stable PDL crossing DE model using this constraint and with the least number of model parameters
possible, so as to determine these parameters with less uncertainties by fitting with the
observational data. While within the scope of two k-fields the stability constraint leads to
complicated field solutions, we construct a reasonably simpler model involving three k-fields for
which stable PDL crossing is possible.

We constrain the parameters of the model using the most recent data-set released by the Supernova
Cosmology Project (SCP) team \cite{kowal}, viz., the ``Union" compilation, which consists of $414$
SNIa reduced to $307$ usable data points after various selection cuts. Marginalizing over the Hubble
constant $H_0$, we find that the best fit value of the present matter density parameter $\O_{0m}
\simeq 0.265$, whereas the best fit DE EoS parameter $\wx$ crosses the PDL at red-shift $z = z_c
\sim 0.2$, fairly in agreement with $z_c$ found in many model-independent DE parameterizations
\cite{pdlc}. The minimized value of the $\chi^2$ statistic is found to be about $310.4$, which shows
an improvement of $\D\chi^2 \sim - 2.7$ over the minimized $\chi^2$ for the $\L$CDM model found
with the Union data-set \cite{rubin}. The minimized $\chi^2$ per degree of freedom for the model
discussed in this paper is $1.011$, which is also marginally better than that found in \cite{davis}
for the $\L$CDM model using the older SNIa data-set complied by Riess {\it et al} \cite{riess2}.
Additionally, we see that the best fit value of the DE density almost tracks the matter density
over sufficiently longer red-shift range in the past until becoming dominant after $z \simeq 0.4$.
Thus the model provides a possible resolution of the coincidence problem. Extrapolation of the best
fit $\wx$ to future epochs shows a tendency of $\wx$ to turn back towards $-1$ after reaching a
minimum value. Thus the phantom regime appears to be transient and future singularities, such as the
Big Rip, may be avoided.

The present model also differs from the usual quintom model \cite{quintom} implemented using two (or
possibly more) scalar fields, (at least) one of which has a negative definite kinetic energy density
(i.e., phantom), and as such the EoS parameter is less than $-1$, whereas the other field(s) has(have)
positive definite kinetic energy (quintessence-type) and EoS parameter(s) greater than $-1$. The
negative energy field comes to dominate at late times and is responsible for the PDL crossing. In the
three-field k-essence model which we study, two of the k-fields have positive definite energy density,
whereas the energy density of the third k-field is negative for a sufficiently longer time in the past
and present, but may become positive in distant future, as found by extrapolating the best fit model
parameters to future (negative) red-shifts. In that sense the third k-field may not be regarded as
equivalent to an eternal phantom, and quantum mechanical instabilities may be avoided. None of the
fields have EoS parameter less than $-1$, unlike quintom models. Moreover, the negative energy k-field
is not responsible for the super-acceleration ($\wx < -1$), it is a positive energy k-field that
causes the PDL crossing, again in contrast with what happens in quintom models.

This paper is organized as follows: in sec. \ref{sec:gf} we describe the general framework of
multiple k-essence DE models assuming that there are no mutual interactions between the
individual k-fields. In sec. \ref{sec:kks} we consider the simplistic case of purely kinetic
multiple k-essence and show that only for more than one dynamical k-fields the PDL crossing is
possible, although the field configuration is severely constrained by the stability criterion.
We work out this constraint and choose the value of the speed of propagation $\cx$ of the DE
density perturbations to be a constant, in order to have a stable PDL crossing multiple kinetic
k-essence DE model. In sec. \ref{sec:model} we construct a specific multiple k-field model for
which $\cx$ is constant, and $\wx$ crosses the PDL in recent past. In sec. \ref{sec:modelfit}
we fit this model with the $307$ Union SNIa data \cite{kowal}, to reconstruct phenomenologically
the multiple k-essence Lagrangian. We conclude with a summary and open questions in sec.
\ref{sec:concl}. In the Appendix, we work out the equations of scalar-type cosmological
perturbations for a multiple k-essence field configuration so as to find the expression for
the propagation speed $\cx$.

\section{General Formalism        \la{sec:gf}}

Let us consider the following action in $(3+1)$ dimensions:
\be \la{gen-action}
\fl S = \int d^4 x \sq{-g} \le[\fr R {2 \k^2} + P ({\mathbf \F}, {\mathbf X}) +
{\mathcal L}_m \ri]
\, , \quad P ({\mathbf \F}, {\mathbf X}) = \sum_{I=0}^{N-1} \pI (\fI, \xI) \, ,
\ee
where $\k^2 = 8 \p G$ is the gravitational coupling constant, ${\mathcal L}_m$ is the
Lagrangian density for perfect fluid matter, ${\mathbf \F} = (\f_{_0}, \cdots, \f_{_{N-1}})$
are a set of $N$ mutually non-interacting scalar fields, and ${\mathbf X} = (X_{_0}, \cdots,
X_{_{N-1}})$ where $\xI$ stands for the kinetic term of the scalar field $\fI$:
\be \la{k-term}
\xI ~=~ -~ \fr 1 2~ g^{\m\n}~ \pa_\m \fI~ \pa_\n \fI \quad, \qquad ({\rm no~ sum~
over}~~ I = 0, \cdots, N-1) \, . \qquad
\ee
$\pI (\fI, \xI)$ is the k-essence Lagrangian density for the field $\fI$.

We assume the space-time structure to be described by a spatially flat Robertson-Walker
(RW) line-element
\be \la{RW}
ds^2 = - dt^2 + a^2 (t) \le[dr^2 + r^2 \le(d\vth^2 + \sin^2 \vth d\vph^2\ri)\ri]
\ee
where $a (t)$ is the scale factor which is normalized to unity at the present epoch
$t = t_0$. In such a space-time, the kinetic term reduces to $\xI = \dfI^2/2$,
where the over-dot stands for $d/dt$. The field equations are the Friedmann equations
given by
\be \la{FRW-eom}
\fl H^2 \equiv \le(\fr{\dot{a}}a\ri)^2 = \fr{\k^2} 3 \le(\r_m + \rx\ri) \quad , \qquad
\dot{H} \equiv \fr{\ddot{a}}a - \fr{\dot{a}^2}{a^2} = - \fr{\k^2}2 \le[\r_m + \rx +
p_m + \px\ri] \, ,
\ee
where $H = \dot{a}/a$ is the Hubble expansion rate, $\r_m$ and $p_m$ are the perfect
fluid energy density and pressure respectively, $\rx$ and $\px$ are the k-essence
energy density and pressure respectively, given by
\be \la{k-ed-pr}
\fl \rx ~=~ \sum_{I=0}^{N-1} \rI (\fI, \xI) ~=~ \sum_{I=0}^{N-1} \le(2 \xI \fr {\pa \pI}
{\pa \xI} ~-~ \pI\ri) \quad, \qquad \px ~=~ \sum_{I=0}^{N-1} \pI (\fI, \xI) \, .
\ee
Considering the perfect fluid matter in the form of pressure-less dust ($p_m = 0$),
the field equations integrate to give the matter energy density as $\r_m = \r_{0m}/a^3$,
where $\r_{0m}$ is the present value of $\r_m$. From Eq.(\ref{FRW-eom}) one obtains the
expression for the normalized Hubble rate:
\be \la{norm-hub}
{\tilde H} ~\equiv~ \fr H {H_0} ~=~ \le[\fr{\rx}{\r_{0c}} ~+~ \fr{\O_{0m}}
{a^3}\ri]^{1/2} \, ,
\ee
where $H_0$ is the present value of the Hubble parameter $H$ and $\r_{0c} = 3
H_0^2/\k^2$, $\O_{0m} = \r_{0m}/\r_{0c}$ are the present values of the critical
density and matter density parameter respectively.

The equations of motion of the scalar fields ${\mathbf \F}  = (\f_{_0}, \cdots,
\f_{_{N-1}})$ are given as
\be \la{sf-eom}
\fr d{dt} \le(a^3 \dfI \fr {\pa \pI}{\pa \xI}\ri) = a^3 \fr{\pa \pI}{\pa \fI}
\qquad (\mbox{for all} ~~I) \, ,
\ee
and the mutually non-interacting individual field components satisfy the continuity
equation:
\be \la{cont} \drI = 3 H \le(1 + \wI\ri) \pI \qquad (\mbox{for
all} ~~I) \, , \ee
where $\wI = \pI/\rI$ are the EoS parameters for the individual
fields $\fI$. The EoS parameter $\wx$ for the k-essence DE is thus
given by
\be \la{de-eos} \wx ~=~ \fr{\px}{\rx} ~=~ \sum_{I=0}^{N-1} \fr{\wI
\rI}{\rx} ~=~ - 1 - \fr{\drx}{3 H \rx} \, . \ee
Using  Eqs. (\ref{FRW-eom}) and (\ref{norm-hub})
one can express the total EOS parameter $w$ in terms of $\wx$:
\be \la{tot-eos}
w ~=~ \fr{\px}{\r_m + \rx} ~=~ \wx \le(1 - \fr{\O_{0m}}{{\tilde H}^2 a^3}\ri) \, .
\ee
The deceleration parameter $q$ is given by
\be \la{q}
q ~=~ - \fr{\ddot a}{a H^2} ~=~ \fr{1 + 3 w} 2 \, .
\ee

For an accelerated expansion of the universe ($q < 0$) the total EOS parameter $w$ should
be less than $- 1/3$, and as can be observed from Eq. (\ref{tot-eos}), $\wx$ is further
less than $-1/3$. Although the observations favour the value of $\wx$ to be very close to
that for a cosmological constant (i.e., $-1$) at present and in recent past, extensive
studies show that better fits with observational data require a dynamical DE with $\wx$
substantially greater than $-1$ at relatively higher red-shifts ($z = a^{-1} - 1 \gtrsim
1$). Moreover, for consistent statistical data fits it is not desirable to always have a
theoretical restriction such as $\wx \geq -1$. In other words, the dynamical DE models
that allow the PDL crossing are most suited for analyzing the observational data so as
to determine the cosmic evolution history over the entire red-shift range that is probed.
While such a crossing in minimally coupled scalar field DE models is associated with violent
cosmological instabilities, one has to ensure that the instabilities never occur in the
multiple k-essence DE models. For such models the expression of the square of the speed
of propagation of curvature perturbations is found to be (see the Appendix for detailed
derivation)
\be \la{ceff}
\cx^2 = (\rx + \px) \le[ \sum_{I=0}^{N-1} \fr{\rI + \pI}{\ctI^2} \ri]^{-1} \quad,
\qquad \mbox{where} ~~ \ctI^2 = \fr{\pa \pI/\pa \xI}{\pa \rI/\pa \xI} \, .
\ee
In the case of a single k-field ($I=0$, say) the above Eq. (\ref{ceff}) reduces to the
well-known result $\cx^2 = \tilde{c}_{_0}^2 = \fr{\pa P_{_0}}{\pa X_{_0}} {\big /}
\fr{\pa \r_{_0}}{\pa X_{_0}}$ \cite{gm}. The value of $\cx^2$ must be non-negative,
for the stability of a model, and always be less than unity (in natural units) so that
the DE perturbations do not travel faster than light.

In the next section we examine the plausibility of a stable PDL crossing in the multiple
k-essence scenario, considering for simplicity the k-field Lagrangian to be purely
kinetic in nature.

\section{Purely kinetic multiple k-essence        \la{sec:kks}}

When the k-essence Lagrangian is purely a function of the kinetic terms ${\mathbf X} =
(X_{_0}, \cdots, X_{_{N-1}})$, i.e., $\pI = \pI (\xI), ~(I = 0 \cdots, N-1)$, the
equations of motion (\ref{sf-eom}) for the scalar fields ${\mathbf \F} = (\f_{_0},
\cdots, \f_{_{N-1}})$ can be integrated at once to give
\be \la{P-deriv}
a^3 \dfI \fr{d \pI}{d \xI} = \mbox{constant} ~~ (k_{_I}) \quad, \qquad \mbox{for all} ~~I \, .
\ee
However, remembering that $\xI = \dfI^2/2$ in Robertson-Walker space-time, one may observe
that the expressions (\ref{k-ed-pr}), (\ref{de-eos}) and (\ref{ceff}), for $\rx$, $\wx$ and
$\cx^2$ respectively, remain invariant under a field redefinition $\fI \rightarrow k_{_I} \fI$.
Hence the constants $k_{_I}$ can effectively be set to unity, for all $I$, and the above equation
can be rewritten as
\be \la{P-deriv1}
\sq{2 \xI} \fr{d \pI}{d \xI} = \fr 1 {a^3} \quad, \qquad \mbox{for all} ~~I \, .
\ee

Now, using the definitions (\ref{k-ed-pr}) of $\rx$ and $\rI$, we can express the DE EoS
parameter $\wx$, Eq. (\ref{de-eos}), and the square of the propagation speed of perturbations
$\cx^2$, Eq. (\ref{ceff}), as
\be \la{wx-cxsq}
\fl \wx = - 1 + \fr 1 {\rx} \sum_{I=0}^{N-1} 2 \xI \fr{d \pI}{d \xI} \quad ,
\quad \cx^2 = \fr{\sum _{I=0}^{N-1} 2 \xI d \pI/d \xI}{\sum _{I=0}^{N-1} \le[2 \xI
d \pI/d \xI + 4 \xI^2 d^2 \pI/d \xI^2\ri]} \, . \quad
\ee
For a transition from $\wx > -1$ to $\wx < -1$, the quantity $\sum 2 \xI d \pI/d \xI$ needs
to vanish at the PDL cross-over scale factor $a = a_c$.

In the case of a single scalar field $\fI = \f_{_0}$, say, and $\pI (\xI) = P_{_0} (X_{_0})$,
the PDL crossing implies that $X_{_0} (dP_{_0}/dX_{_0}) \vline_{a = a_c} = 0$, i.e., either
$X_{_0} (a=a_c)$ or $dP_{_0}/dX_{_0} \vline_{a = a_c}$ or both must vanish. This is clearly not
possible for any finite value of $a$ since from Eq.(\ref{P-deriv1}) we have $\sq{2 X_{_0}}
(dP_{0}/dX_{_0}) = a^{-3}$.

For more than one k-fields the PDL crossing is in principle possible whenever the quantity
$\sum 2 \xI (d\pI/d\xI) = 0$ at $a = a_c$, i.e., ~$\sum \sq{2 \xI} \vline_{a = a_c} = 0$, by Eq.
(\ref{P-deriv1}). However, from Eq. (\ref{wx-cxsq}) we find that the change of sign of $\sum 2
\xI (d\pI/d\xI)$ as the $\wx = -1$ line is crossed, implies a change of sign of sign of
$\cx^2$ as well, unless the numerator and denominator of $\cx^2$ exactly cancel each other
to give a value which never flips sign. For example, if $\pI$ is a function of some numerical
power of $\xI$, i.e., $\pI = \xI^{\nuI}$ (for all $I$), as has been considered in ref.
\cite{chim}, there is always a flip of sign of $\cx^2$ associated with the PDL crossing.
Therefore unless one chooses $\cx^2$ to be such a function of $a$ which does not change sign
for any value of $a$, instabilities are bound to be there even in multiple k-essence DE models.
Moreover, we must have $\cx^2 \leq 1$ in order that the DE perturbations do not travel faster
than light. The simplest possible choice of $\cx^2$, appropriate for a stable and non-super-luminal
DE model is
\be \la{cxsq-const}
\cx^2 ~=~ n ~(\mbox{constant}) \quad, \qquad 0 \leq n \leq 1 \, .
\ee
Of course, the limiting case $n = 0$ implies a non-propagating DE (for example, due to a
cosmological constant). For dynamic DE models we have $0 < n \leq 1$.

Let us now denote
\be \la{y-def}
\yI = \sq{2 \xI} = \dfI ~~\mbox{in RW space-time} \, .
\ee
Eq. (\ref{P-deriv1}) reduces to
\be \la{P-deriv-y}
\fr{d\pI}{d\yI} = \fr 1 {a^3} \qquad \mbox{for all}~ I \, ,
\ee
and the DE energy density takes the form
\be \la{ed-pr}
\rx = \sum_{I=0}^{N-1} \rI \quad, \qquad \rI = \fr{\yI}{a^3} - \pI \, .
\ee
From these equations we find
\be \la{yi}
\yI (a) = - \fr{a^4 \rI' (a)} 3 \qquad \mbox{for all}~ I \, .
\ee
where prime $\{'\}$ denotes $d/da$. Eqs. (\ref{wx-cxsq}) for $\wx$ and $\cx^2$ reduce to
\be \la{wx-cxsq1}
\fl \wx = - 1 + \fr {y (a)} {a^3 \rx} \quad, \quad \cx^2 = - \fr {a y (a)} 3
\le[\sum _{I=0}^{N-1} \yI^2/\yI'\ri]^{-1} \, , \quad  \mbox{where} ~~ y (a) =
\sum _{I=0}^{N-1} \yI (a) \, . \quad
\ee
We rewrite the expression for $\cx^2$ in the form
\be \la{cx-eq}
3~ \cx^2 \fr{\yI^2}{\yI'} ~+~ a \yI (a) ~=~ -~ \etaI (a) \, ,
\ee
where the arbitrary functions $\etaI (a)$ satisfy the relation $\sum \etaI (a) = 0$.
With the choice $\cx^2 = n$ (constant), we therefore have
\be \la{stab-eq}
\fr{\yI'}{\yI} \le[1 + \fr{\etaI (a)}{a \yI (a)}\ri] = -~ \fr {3 n} a \quad , \qquad
\sum _{I=0}^{N-1} \etaI (a) = 0 \, .
\ee

While from Eq. (\ref{wx-cxsq1}) we see that the PDL crossing requires $y = \sum \yI = 0$
at $a = a_c$, a stable k-field configuration is the one that solves the above equation
(\ref{stab-eq}). In the next section we construct a cosmologically stable PDL crossing
DE model by suitably choosing a specific field configuration and a reasonably simplified
ansatz to solve for $y$, $\wx$ and $\rx$. We fit the solutions to the latest SNIa Union data
\cite{kowal} in the subsequent section, show the plausible tracking of the dark energy
density to the matter density in the past, avoidance of future singularities, and finally
reconstruct the multiple kinetic k-essence Lagrangian.

\section{Cosmologically stable PDL crossing multiple kinetic k-essence model  \la{sec:model}}

\subsection{Two k-field configuration    \la{sec:2f-model}}

Let us first consider a field configuration involving only two scalar fields:
\be \la{2f}
I = 0, 1 \quad ; \qquad {\mathbf \F} = (\f_{_0},\f_{_1}) \quad ; \qquad \eta_{_0} (a) =
- \eta_{_1} (a) = \eta (a) \, ,~~\mbox{say} \, .
\ee
From Eq. (\ref{stab-eq}) we have
\be \la{y-eqs}
\fr{\ya'}{\ya} \le(1 + \fr{\eta}{a \ya}\ri) =~ \fr{\yb'}{\yb} \le(1 - \fr{\eta}{a \yb}\ri)
= -~ \fr{3 n} a \, .
\ee
Let us assume a power-law solution for $\yb$ as
\be \la{y1}
\yb ~=~ \ab~ a^s \quad , \qquad \ab, s = \mbox{constants} \, .
\ee
Plugging this in Eq. (\ref{yi}) and integrating, and using Eq. (\ref{ed-pr}), we get the
expressions for the energy density and the pressure corresponding to the field $\f_{_1}$
as
\be \la{rho1-P1}
\fl \r_{_1} = \left\{
\begin{array}{lll}
\le(\fr 3 {3 - s}\ri) \ab a^{s-3} + \l_{_1} \, & & (s \neq 3) \\
- 3 \ab \ln a + \l_{_1} & &  (s = 3)
\end{array}
\right.
\, , \,
P_{_1} = \left\{
\begin{array}{lll}
- \le(\fr s {3 - s}\ri) \ab a^{s-3} - \l_{_1} \, & & (s \neq 3) \\
\ab a^{s-3} - 3 \ab \ln a - \l_{_1} & & (s = 3)
\end{array}
\right.
\ee
where $\l_{_1}$ is a constant.

Substituting Eq. (\ref{y1}) in Eq. (\ref{y-eqs}) and eliminating $\eta(a)$ we obtain
\be \la{alpha-eq}
\le(1 + \fr s {3 n}\ri) \fr{\ya'}{\ya} = \fr{\le(1 + \ya/\yb\ri)'}{1 + \ya/\yb} \, .
\ee
Integrating this we get the algebraic equation
\be \la{alpha-eq1}
\ya ~+~ \yb \le(1 - C \ya^{1 + s/3n}\ri) =~ 0 \, ,
\ee
where $C$ is an integration constant.

We look for the simplest possible solutions of the above equation (\ref{alpha-eq1}),
so that the cosmological quantities $\rx$ and $\wx$ (and hence the normalized Hubble
expansion rate ${\tilde H}$) have reasonably simple forms which we can fit with the
observational data.

Let us first take into account the cases $s = - 3 n$ or $s = 0$ for either of which
Eq. (\ref{alpha-eq1}) is linear in $\yc$. However, for $s = - 3 n$, Eq. (\ref{alpha-eq1})
gives $\ya \propto \yb$, i.e., $y = \ya + \yb \propto a^{- 3 n}$, which can never
change sign for any finite value of $a$. Therefore we rule out $s = - 3 n$ as a possible
choice for a PDL crossing DE model. For $s = 0$, on the other hand, we obtain from Eqs.
(\ref{y1}) and (\ref{alpha-eq1}) that $\ya = \yb/(C \yb - 1) =$ constant, i.e., the
corresponding fields $\f_{_1}$ and $\f_{_2}$ are non-dynamical, and PDL crossing never
happens either. Hence we rule out the choice $s = 0$ as well.

We next consider the case $s = 3 n/2$ for which Eq. (\ref{alpha-eq1}) is quadratic in
$\ya$. Corresponding to $0 \leq n \leq 1$, we have $0 \leq s \leq 3$. The solution of
Eq. (\ref{alpha-eq1}) is given by
\be \la{y0}
\ya ~=~ \aa~ a^{-s} \le[1 \pm \sqrt{1 + \fr{2 \ab a^{2 s}}{\aa}}\ri] \, ,
\ee
where we denote $\aa = 1/(2 C \ab)$. From Eqs. (\ref{y1}) and (\ref{y0}) we see once
again that the expression for $y = \ya + \yb$ cannot change sign, i.e., the PDL crossing
is not possible.

We are therefore left with the choices of the parameter $s$ which render Eq. (\ref{alpha-eq1})
to be cubic, quartic or even higher order algebraic equation in $\ya$\footnote{There are also
the choices $s = - 3n/2$ and $s = -6n$ for which Eq. (\ref{alpha-eq1}) is quadratic in
$\sqrt{\ya}$ and $\ya$ respectively. However, the solution for $\ya$ corresponding to these
choices is not qualitatively different from that corresponding to $s = 3n$ shown above. The
PDL crossing can never take place either.}. The solutions for $\ya$ are however very complicated
and leads to messy expressions for the corresponding energy density. An inspection of Eq.
(\ref{alpha-eq1}) also reveals that the choices of $s$ for which Eq. (\ref{alpha-eq1}) is even
order (quadratic, quartic, etc.) in $\ya$, can be ruled out anyway, because $y = \ya + \yb =
C \yb \ya^{1+s/3n}$ can in no way change sign when the power $(1 + s/3n)$ is an even integer.
In a separate work which is in progress \cite{sssd} we examine the cases for which Eq.
(\ref{alpha-eq1}) is cubic or higher order in $\ya$. In the present paper, however, we choose
to work with a simpler alternative by introducing one more k-field, in a way that Eq.
(\ref{alpha-eq1}) remains quadratic in $\ya$, and the above solution (\ref{y0}) could be used.

\subsection{Three k-field configuration    \la{sec:3f-model}}

Let us now consider the following configuration that involves three scalar fields:
\be \la{3f}
\fl I = 0,1,2 \quad ; \qquad {\mathbf \F} = (\f_{_0},\f_{_1},\f_{_2}) \quad ; \qquad \eta_{_0} (a)
= - \eta_{_1} (a) = \eta (a) \, , \quad \eta_{_2} = 0 \, .
\ee
From Eq. (\ref{stab-eq}) we have
\be \la{y3-eqs}
\fr{\ya'}{\ya} \le(1 + \fr{\eta}{a \ya}\ri) =~ \fr{\yb'}{\yb} \le(1 - \fr{\eta}{a \yb}\ri)
~=~ \fr{\yc'}{\yc}  ~=~ -~ \fr{3 n} a \, .
\ee
The solution for $\yc$ is
\be \la{y2}
\yc ~=~ \ac~ a^{- 3 n} \quad , \qquad \ac = \mbox{constant} \, .
\ee
whence from Eqs. (\ref{ed-pr}) and (\ref{yi}) we find the energy density and the pressure
corresponding to the field $\f_{_2}$ as
\be \la{rho2-P2}
\r_{_2} ~=~ \fr{\ac}{n + 1} a^{- 3 (n + 1)} + \l_{_2} \quad , \qquad
P_{_2} ~=~ \fr{n \ac}{n + 1} a^{- 3 (n + 1)} - \l_{_2} \, ,
\ee
where $\l_{_2}$ is a constant.

Once again, we assume the power-law solution (\ref{y1}) for $\yb$, so that the energy density
and the pressure corresponding to the field $\f_{_1}$ are given by (\ref{rho1-P1}), and $\ya$
is related to $\yb$ through the relation (\ref{alpha-eq1}). Similar to the case of two k-fields,
it is easy to show that the PDL crossing never happens for three k-fields as well, if the
parameter $s = - 3n$ or $0$ so that Eq. (\ref{alpha-eq1}) is linear in $\ya$. We therefore
resort to the choice $s = 3n$ for which Eq. (\ref{alpha-eq1}) is quadratic in $\ya$ and the
solution for $\ya$ is given by Eq. (\ref{y0}). Using Eqs. (\ref{y1}), (\ref{y2}) and (\ref{y0})
we obtain
\be \la{y-total}
y ~=~ \ya + \yb + \yc ~=~ \le(\aa + \ac\ri) a^{-s} ~+~ \ab a^s ~\pm~ \aa a^{-s}
\sqrt{1 + \fr{2 \ab a^{2 s}}{\aa}} \, . \qquad
\ee
For a PDL crossing at $a = a_c$, $y (a = a_c) = 0$, which implies that the parameters
$\aa, \ab$ and $\ac$ should satisfy the relation
\be \la{rel-alpha}
\le(\ac + \ab a^{2 s}\ri)^2 + 2 \aa \ac = 0 \, .
\ee
Thus a smooth crossing from a $\wx > -1$ regime to a $\wx < -1$ regime is indeed possible
for the choice $s = 3 n$, provided one of the parameters $\aa$ and $\ac$ must be negative
with respect to the other, i.e., $\aa \ac < 0$. However, a negative value of $\ac$ implies
an eternally negative definite kinetic energy density $\r_{_2}$ for the field $\f_{_2}$. A
negative value $\aa$, on the other hand, does not necessarily imply a negative energy
density for the field $\f_{_0}$, which is derived below. We therefore choose to work with
$\aa < 0, \ac > 0$.

Let us now set $s = 1/2$, so that the DE perturbations are slowly propagating with speed
$\cx = \sqrt{n} = 1/\sqrt{6}$, and for which Eq. (\ref{yi}) could be integrated easily. Using
Eqs. (\ref{ed-pr}) and (\ref{yi}) one then obtains the expressions for the energy density
and pressure components due to the field $\f_{_0}$ as
\bea \la{rho0-P0}
\fl \r_{_0} &=& \fr{6 \aa}{7 a^{7/2}} \le[1 \pm \le(1 + \fr{4 \b a} 5 + \fr{8 \b^2 a^2}{15}\ri)
\le(1 - \b a\ri)^{3/2}\ri] + \l_{_0} \, , \nn \\
\fl P_{_0} &=& \fr{\aa}{7 a^{7/2}} \le[1 \pm \le\{1 + \fr{6 \b a} 5 \le(1 + \fr{4 \b a} 3 +
\fr{8 \b^2 a^2} 3\ri)\ri\} \le(1 - \b a\ri)^{1/2}\ri] - \l_{_0} \, ,
\eea
where $\l_{_0}$ is a constant and we denote $\b = - 2 \ab/\aa$. Using Eqs. (\ref{rho1-P1}),
(\ref{rho2-P2}) and (\ref{rho0-P0}) we write the expressions for the total dark energy
density ($\rx = \r_{_0} + \r_{_1} + \r_{_2}$) and total pressure ($\px = P_{_0} + P_{_1} +
P_{_2}$), for the choice $s = 3 n = 1/2$, as
\bea \la{ded-pr}
\fl \rx &=& \fr{6 \le(\aa + \ac\ri)}{7 a^{7/2}} + \fr{6 \ab} {5 a^{5/2}} \pm
\fr{6 \aa}{7 a^{7/2}} \le(1 + \fr{4 \b a} 5 + \fr{8 \b^2 a^2}{15}\ri)\le(1 - \b a\ri)^{3/2}
+ \l \, , \nn\\
\fl \px &=& \fr{\aa + \ac}{7 a^{7/2}} - \fr{\ab}{5 a^{5/2}} \pm
\fr{\aa}{7 a^{7/2}} \le[1 + \fr{6 \b a} 5 \le(1 + \fr{4 \b a} 3 + \fr{8 \b^2 a^2} 3\ri)\ri]
\le(1 - \b a\ri)^{1/2} - \l \, ,
\eea
where $\l = \l_{_0} + \l_{_1} + \l_{_2}$.

Now, from Eq. (\ref{norm-hub}) we express the normalized Hubble rate (${\tilde H} = H/H_0$)
as
\be \la{hub}
\fl {\tilde H}^2 = \fr{\O_{0m}}{a^3} + \fr{6 \le(\Aa + \Ac\ri)}{7 a^{7/2}} + \fr{6 \Ab}
{5 a^{5/2}} \pm \fr{6 \Aa}{7 a^{7/2}} \le(1 + \fr{4 \b a} 5 + \fr{8 \b^2 a^2}{15}\ri)
\le(1 - \b a\ri)^{3/2} + \L \, ,
\ee
where $\Aa = \aa/\r_{0c}, \Ab = \ab/\r_{0c}, \Ac = \ac/\r_{0c}, \b = -2 \ab/\aa = - 2 \Ab/\Aa$
and $\L = \l/\r_{0c}$. Eliminating $\L$ using the condition that at the present epoch $t = t_0$,
$a = 1$ and ${\tilde H} = 1$, we obtain
\bea \la{hub1}
\fl {\tilde H}^2 &=& 1 + \O_{0m} \le(a^{-3} - 1\ri) + \fr{6 \le(\Aa + \Ac\ri)} 7 \le(a^{-7/2} -
1\ri) + \fr{6 \Ab} 5 \le(a^{-5/2} - 1\ri) \nn \\
\fl &\pm& \fr{6 \Aa} 7 \le[a^{-7/2} \le(1 + \fr{4 \b a} 5 + \fr{8 \b^2 a^2}{15}\ri)\le(1 - \b
a\ri)^{3/2} - \le(1 + \fr{4 \b} 5 + \fr{8 \b^2}{15}\ri)\le(1 - \b\ri)^{3/2}\ri] .
\eea

One may observe that the parameter $\b$ should not be greater than $1$ in order that ${\tilde H}^2$
is real, and the model to be valid. Moreover, for positive fractional values of $\b$, for which
the PDL crossing could take place, the range of validity of the model is up to a finite time in
the future at which the scale factor is $a_{max} = 1/\b$. Therefore, the lesser the positive
fractional value of $\b$, the more the model could be extended to future epochs.

In what follows, we fit this model with the observational data so as to determine the parameters
$\Aa, \Ab$ and $\Ac$, and finally reconstruct the Lagrangian components $P_{_0} (\ya), P_{_1} (\yb)$
and $P_{_2} (\yc)$ using the best fit values of the parameters.

\section{Fitting the model with observational data  \la{sec:modelfit}}

We use the most recent SNIa data released by the SN Cosmology Project (SCP) team, viz., the Union
data-set \cite{kowal}. It consists of a total of $414$ SN samples analyzed from various sources,
reduced to $307$ most reliable data points after different outlier rejection cuts and selection
tests. These $307$ Union data includes large samples of SNIa from the older data-sets by Perlmutter
{\it et al}, Tonry {\it et al}, Riess {\it et al}, and others \cite{perl,tonry,riess1,riess2}, SN
Legacy Survey (SNLS) \cite{astier}, as well as the distant SN observed by the Hubble Space Telescope
(HST). The range of data extends up to red-shift $z \sim 1.7$, and the full data can be obtained at
http://supernova.lbl.gov/Union.

These data provide the extinction-corrected distance modulus, given by
\be \la{dist-mod}
\m_{obs} (z_i) ~=~ m_{obs} (z_i) ~-~ M \, ,
\ee
where $m_{obs}$ is the apparent magnitude of the source SN and $M$ is the absolute magnitude.
Theoretically, the apparent magnitude is expressed as
\be \la{app-mag-th}
m_{th} (z_i) ~=~ 5 \log_{10} \le[D_L (z_i)\ri] +~ {\bar M} (M, H_0) \, ,
\ee
where $D_L (z_i)$ is the Hubble free luminosity distance, which for a spatially flat universe is
defined in terms of the present matter density parameter $\O_{0m}$ and the set of theoretical model
parameters ${\mathbf A} = (\Aa, \dots, \An)$ as
\be \la{lum-dist}
D_L (z_i) ~=~ \le(1 + z_i\ri) \int_0^{z_i} \fr{d {\bar z_i}}{{\tilde H} ({\bar z_i}; \O_{0m},
{\mathbf A})} \, .
\ee
${\bar M}$ is the magnitude zero point offset given by
\be \la{zeropt}
{\bar M} (M, H_0) = M + 5 \log_{10} \le[\fr{H_0^{-1}}{Mpc}\ri] + 25 = M - 5 \log_{10} h +
42.38 \, ,
\ee
where $h$ is the Hubble constant $H_0$ in units of $100$ Km s$^{-1}$ Mpc$^{-1}$.

The theoretical distance modulus $\m_{th} (z_i)$ is defined by
\be \la{dist-mod-th}
\m_{th} (z_i) ~=~ m_{th} (z_i) - M ~=~ 5 \log_{10} \le[D_L (z_i)\ri] + \m_0 \, ,
\ee
where
\be \la{nuis}
\m_0 ~=~ 42.38 ~-~ 5 \log_{10} h \, ,
\ee
is a nuisance parameter, independent of the data points, and has to be uniformly marginalized
over (i.e., integrated out).

The likelihood of $\O_{0m}$ and the theoretical model parameters ${\mathbf A} = (\Aa, \dots, \An)$
are determined by minimizing the quantity
\be \la{chi2}
\chi^2 (\m_0; \O_{0m}, {\mathbf A}) ~=~ \sum_{i=1}^{N_{tot}} \fr{\le[\m_{obs} (z_i) -
\m (z_i)\ri]^2}{\s_i^2 (z_i)} \, ,
\ee
where $\s_i (z_i), (i = 1, \dots, N_{tot})$, are the total uncertainties on the data points. For
the Union data-set \cite{kowal} $N_{tot} = 307$.

Now, the marginalization with respect to $\m_0$ can be done by the following two-step procedure
\cite{peri}:

\noindent
1. Expanding $\chi^2$ in terms of $\m_0$ as
\be \la{chi2-expand}
\chi^2 (\m_0; \O_{0m}, {\mathbf A}) ~=~ {\tilde A} (\O_{0m}, {\mathbf A}) ~-~ 2 \m_0 ~
{\tilde B} (\O_{0m}, {\mathbf A}) ~+~ \m_0^2 ~ {\tilde C} \, ,
\ee
where
\bea \la{tildeABC}
{\tilde A} (\O_{0m}, {\mathbf A}) ~=~ \sum_{i=1}^{N_{tot}} \fr{\le[\m_{obs} (z_i) -
\m (z_i; \m_0 = 0)\ri]^2}{\s_i^2 (z_i)} \, , \nn \\
{\tilde B} (\O_{0m}, {\mathbf A}) ~=~ \sum_{i=1}^{N_{tot}} \fr{\m_{obs} (z_i) -
\m (z_i; \m_0 = 0)}{\s_i^2 (z_i)} \, , \quad
{\tilde C} = \sum_{i=1}^{N_{tot}} \fr 1 {\s_i^2 (z_i)} \, ,
\eea
one finds that Eq. (\ref{chi2-expand}) is minimum for $\m_0 = {\tilde B}/{\tilde C}$.

\noindent
2. Substituting this value of $\m_0$ in Eq. (\ref{chi2-expand}) we obtain
\be \la{tchi2}
\fl {\tilde \chi}^2 (\O_{0m}, {\mathbf A}) ~\equiv~ \chi^2 \le(\m_0=\fr{\tilde B}{\tilde C};
\O_{0m}, {\mathbf A}\ri) ~=~ {\tilde A} (\O_{0m}, {\mathbf A}) ~-~ \fr 2 {\tilde C} {\tilde B}^2
(\O_{0m}, {\mathbf A}) \, .
\ee
Now instead of minimizing $\chi^2$ in Eq. (\ref{chi2}), one can minimize ${\tilde \chi}^2$ in
Eq. (\ref{tchi2}) with respect to $\O_{0m}$ and ${\mathbf A} = (\Aa, \dots, \An)$, since both
the minimized values would be the same.

In the case of the three k-field ($N=3$) model discussed above, corresponding to the choice $s
= 3 n = 1/2$, we find good fits to the SNIa data for positive fractional values of $\b = - 2
\Ab/\Aa$, and with the negative sign of square roots appearing in the expression for ${\tilde H}^2$
in Eq. (\ref{hub1}). We fit the observational data for two parametric choices of $\b = 0.5$ and
$0.25$, for which the model remains valid up to sufficiently large future red-shifts $z = - 0.5$
and $ - 0.75$ respectively.

\vskip 0.15in
\noindent
\underline{$\b = 0.5$:}~ With this choice we get the minimized value of $\chi^2 = 310.411$
for the $307$ Union data-set \cite{kowal}, and the best fit values $\O_{0m} = 0.26662, \Ab
= 1.46422, \Ac = 0.159546$. The minimized $\chi^2$ is better by $\D\chi^2 = -2.69$ from that
found with the Union data for the two-parameter flat $\L$CDM model \cite{kowal,rubin}. The
minimized $\chi^2$ per degree of freedom (dof) is also improved by $\D (\chi^2/dof) = -
0.0019$ from that found in ref. \cite{davis} by fitting the flat $\L$CDM model with the older
SNIa gold data-set \cite{riess2}.

\vskip 0.15in
\noindent
\underline{$\b = 0.25$:}~ With this choice we get the minimized $\chi^2 = 310.385$, which is
improved further by $\D\chi^2 = -0.26$ than that obtained for the previous choice ($\b = 0.5$).
The best fit values of the present matter density parameter and the model parameters are obtained
as: $\O_{0m} = 0.26103, \Ab = 3.58355, \Ac = 0.174564$.

\vskip 0.2in
%
\begin{figure}[!htb]
\begin{center}
\includegraphics[width=15cm,height=10cm]{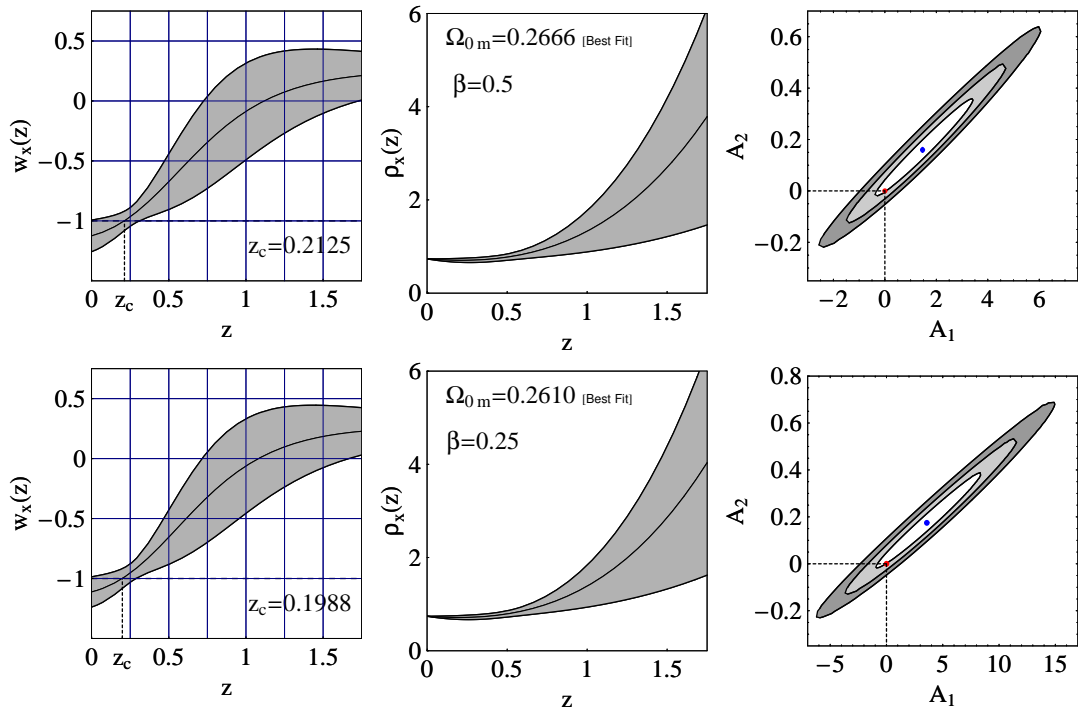}
\end{center}
\caption{\small Evolution of $\wx (z)$ and $\rx (z)$ (best fit with the Union data, plus $1\s$
errors) for $0 \leq z \leq 1.75$, as well as the $1\s$ (white), $2\s$ (light grey) and $3\s$
(dark grey) contours in $\Ab-\Ac$ parameter space, for the choices $\b = 0.5$ (upper panels)
and $\b = 0.25$ (lower panels). The point $z_c$ denotes the value of the red-shift at which
the PDL ($\wx = -1$) is crossed. The best fit points for both the choices are shown by the dots
at the middle of the $1\s$ $\Ab-\Ac$ contours. The cosmological constant, which corresponds to
$\Ab = \Ac = 0$, is found to lie on edge of $1\s$ $\Ab-\Ac$ contour for both the choices.}
\la{fig1}
\end{figure}

The evolution of $\wx (z)$ and $\rx (z)$ (in units of $\r_{0c}$), alongwith the corresponding
$1\s$ errors, for $0 \leq z \leq 1.75$, as well as the $1\s, 2\s$ and $3\s$ $\Ab-\Ac$ contours
are shown in Fig. \ref{fig1}, for both the choices $\b = 0.5$ and $0.25$. One may also have the
$\O_{0m}-\Ab$ and $\O_{0m}-\Ac$ contour plots for either choice, however for brevity we have
suppressed those plots. The cases $\b = 0.5$ and $\b = 0.25$ are potentially not much distinct.
The $\wx (z)$ vs $z$ plots show that the value $z = z_c$ at which the best fit $\wx$ crosses $-1$
is $0.2125$ for $\b = 0.5$ and $0.1988$ for $\b = 0.25$. Both these values of $z_c$ are fairly in
agreement with those obtained for popular model-independent ansatze for $\wx (z)$ \cite{cpl},
$H (z)$ \cite{alam}, $\px (z)$ \cite{visser}, $q (z)$ \cite{gong}, etc. The cosmological constant
(corresponding to $\Ab = \Ac = 0$), is found to be about $1\s$ away from the best fit result for
both the choices $\b = 0.5$ and $\b = 0.25$.

%
\begin{figure}[!htb]
\begin{center}
\includegraphics[width=15cm,height=7cm]{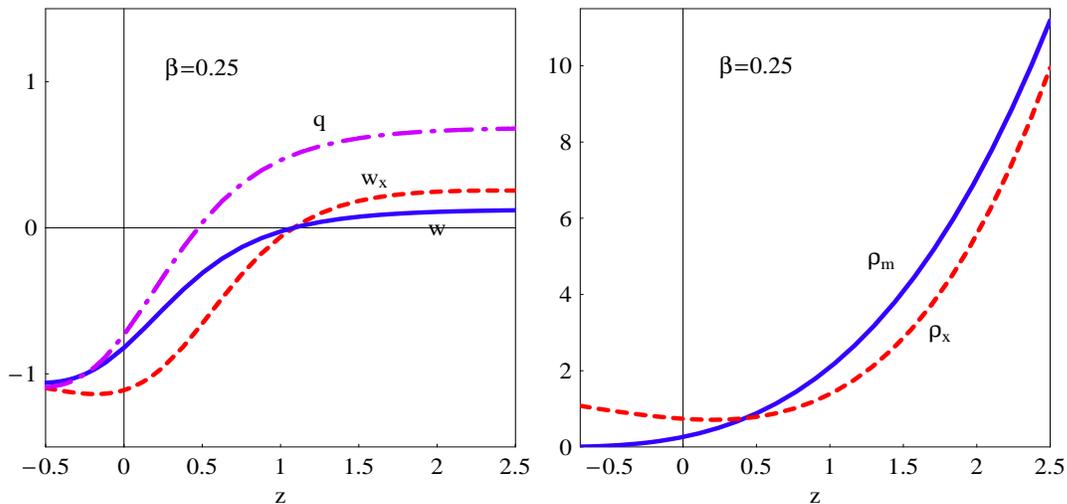}
\end{center}
\caption{\small The left panel shows the extrapolations of $\wx (z), w (z)$ and $q$ (best fit with
the Union data) to the red-shift range $- 0.5 \leq z \leq 2.5$. The DE EoS parameter $\wx$ shows a
tendency to back from a minimum value in future towards the cosmological constant value $-1$. The
right panel shows a nearly tracking nature of the dark energy density $\rx$ (extrapolated best fit)
relative to the matter density $\r_m$ for high red-shifts $z \gtrsim 1$. Both the plots correspond
to the choice $\b = 0.25$.}
\la{fig2}
\end{figure}
%

%
\begin{figure}[!htb]
\begin{center}
\includegraphics[width=15cm,height=10cm]{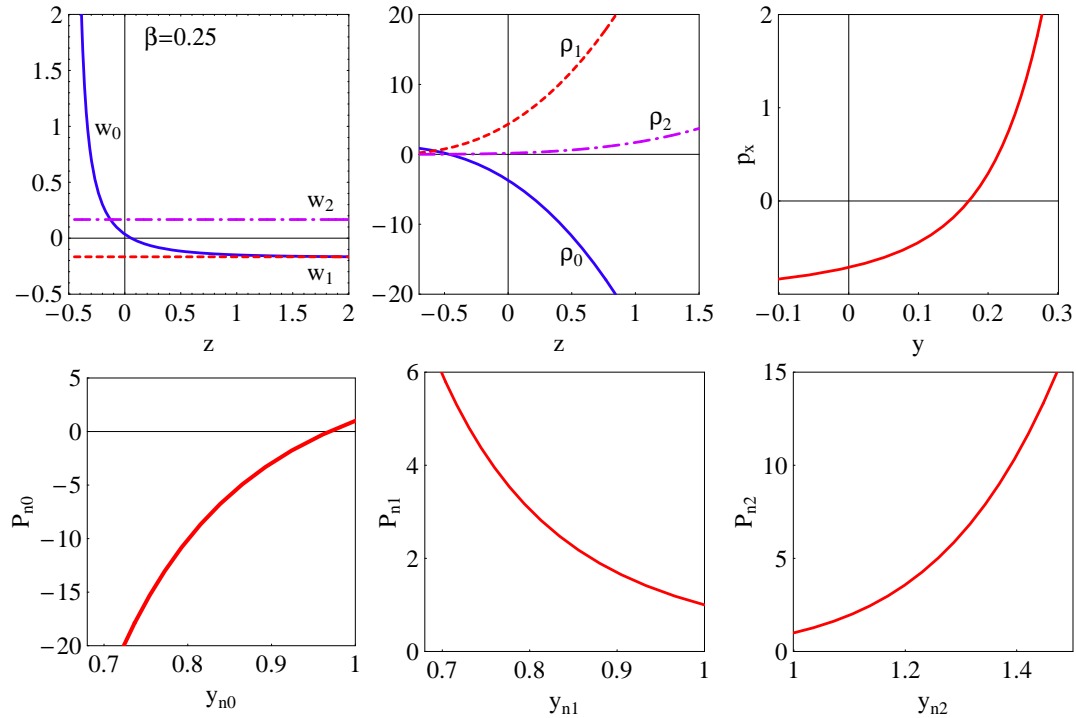}
\end{center}
\caption{\small The upper left panel shows the evolution of the individual EoS parameters $w_{_0} (z),
w_{_1}(z), w_{_2}(z)$ (extrapolated best fit with the Union data) for $- 0.5 \leq z \leq 2$. Whereas
$w_{_1}$ and $w_{_2}$ are negative and positive constants respectively, $w_{_0}$ changes from a
negative to a positive value in very recent past. The upper middle panel shows the evolution of the
individual energy densities $\r_{_0}(z), \r_{_1}(z), \r_{_2}(z)$ (extrapolated best fit with the
Union data) for $- 0.7 \leq z \leq 1.5$. While $\r_{_1}$ and $\r_{_2}$ are positive definite, $\r_{_0}$
changes from a negative to a positive value at a distant future epoch ($z \approx -0.5$). The upper
right panel is a parametric plot (for $0 \leq z \leq 1.75$) showing the variation of the k-essence
Lagrangian $\px = P$ (reconstructed using the best fit model parameters) with $y = \ya + \yb + \yc$.
The lower panels show the parametric plots (for $0 \leq z \leq 1.75$) of the reconstructed normalized
Lagrangian components $P_{n_I} = P_{_I}/P_{_I}^{(0)}$ versus the normalized field gradients $y_{n_I}
= \yI/\yI^{(0)}$ for $I = 0, 1, 2$. All the plots correspond to the choice $\b = 0.25$.}
\la{fig3}
\end{figure}

The left panel of Fig. \ref{fig2} shows the variations of the DE EoS parameter $\wx (z)$, the total
EoS parameter $w (z)$, Eq. (\ref{tot-eos}), and the deceleration parameter $q (z)$, Eq. (\ref{q}),
for the best fit parameters $\O_{0m}, \Ab, \Ac$, extrapolated to the red-shift range $-0.5 \leq z \leq
2.5$. The transition from the decelerating ($q > 0$) phase to the accelerating ($q < 0$) phase occurs at
$z \sim 0.45$ and the values of both $w$ and $q$ are greater than $-1$ at present ($z = 0$). Whereas
the present value of $\wx < -1$, there is a tendency that $\wx$ would turn back from a  minimum
value towards the cosmological constant value ($= -1$) in distant future ($z \lesssim -0.5$). Thus
the phantom ($\wx < -1$) phase may be transient, so that future singularities, such as the Big Rip,
may be avoided. The right panel of Fig. \ref{fig2} shows the variations of the matter density $\r_m$
and the extrapolated best fit DE density $\rx$ (both in units of the present critical density $\r_{0c}$)
with $z$, for $-0.7 \leq z \leq 2.5$. The DE density is found to almost track the matter density for higher
red-shifts, until equalling it at $z \sim 0.4$ and dominant thereafter\footnote{For very high red-shifts
($z \gtrsim 10$) we find that extrapolated $\rx$ exceeds $\r_m$ once again and the DE becomes more and
more dominant as we go further and further in the past. However, the best fit model parameters, which
give rise to such a variation of $\rx (z)$, are found by fitting the Union data available only up to
$z \sim 1.7$. Therefore the extrapolation of $\rx (z)$ to such high red-shifts, using these parametric
values, cannot be trusted with much confidence after all.}. Thus the model apparently resolves the
coincidence problem, side by side providing a PDL crossing DE. The extrapolation of the best fit $\rx$
to future epochs also show that the DE density, though dominant, does not run away to very high values,
causing Big Rip-like singularities. Both the plots of Fig. \ref{fig2} correspond to the choice $\b =
0.25$, for which the model fits better with the Union data, as mentioned above.

All the plots of Fig. \ref{fig3} also correspond to the choice $\b = 0.25$. The upper left panel of
Fig. \ref{fig3} shows the extrapolated best fit variations of the EoS parameters $\wI = \pI/\rI$
corresponding to the individual fields $\fI$ ($I = 0,1,2$) with red-shift $z$, for $-0.5 \leq z \leq
2.5$. While $w_{_1} = -n = -1/6$ and $w_{_2} = n = 1/6$ are negative and positive constants equal in
magnitude, the EoS parameter $w_{_0}$ increases from a negative constant value (equal to $w_{_1} =
-1/6$) in distant past, crosses zero in very recent past ($z \simeq 0.05$) and then goes on increasing
with positive values. The upper middle panel of Fig. \ref{fig3} shows the extrapolated best fit
variations of the individual field energy densities $\rI$ with $z$, for $-0.7 \leq z \leq 1.5$. $\r_{_1}$
and $\r_{_2}$ are positive definite and decreases with time (i.e., with decreasing $z$). $\r_{_0}$ is
negative in the past and at present, but increases with time, and may become positive in distant future
($z \lesssim -0.5$), as the extrapolations show. In magnitude the value of $\r_{_2}$ is found to be very
small compared to $\r_{_0}$ and $\r_{_1}$ for all the red-shift range shown in the figure. $\r_{_0}$ and
$\r_{_1}$, on the other hand, are nearly equal in magnitude for $0 \leq z \leq 0.5$.

Now, the effective DE EoS parameter $\wx$ is given by
\be \la{DE-EoS}
\wx ~=~ \fr{w_{_0} \r_{_0} ~+~ w_{_1} \r_{_1} ~+~ w_{_2} \r_{_2}}{\rx} \quad , \qquad
\rx = \r_{_0} + \r_{_1} + \r_{_2} \, .
\ee
Since $\rx$ is always positive, the negative definite and positive definite contributors to $\wx$ are
$\f_{_1}$ (for which $w_{_1} < 0, \r_{_1} > 0$) and $\f_{_2}$ (for which $w_{_2} > 0, \r_{_2} > 0$)
respectively. The field $\f_{_0}$, on the other hand, gives positive contribution ($w_{_0} \r_{_0} >
0$) to $\wx$ right from the past up to $z \simeq 0.05$ and negative contribution ($w_{_0} \r_{_0} < 0$)
thereafter. However this field $\f_{_0}$, though has a negative energy density (phantom-like), could
not have led to the PDL crossing at $z \simeq 0.2$, as its negative contribution to $\wx$ begins only
after $z \simeq 0.05$. It is therefore the (non-phantom-like) field $\f_{_1}$, with positive definite
energy density $\r_{_1}$, which is effectively responsible for the PDL crossing at $z \simeq 0.2$. This
is in contrast with the usual {\it quintom} mechanism \cite{quintom} where the negative energy density
(phantom) field leads to the PDL crossing at late times. Moreover, none of the three fields above has
EoS parameter $\wI$ less than $-1$, as in quintom models.

We reconstruct the multiple kinetic k-essence Lagrangian $P = \px$ using the best fit values of the
parameters $\O_{0m}, \Ab, \Ac$ for the choice $\b = 0.25$. The parametric plot of $\px$ against the total
temporal field gradient $y = \ya + \yb + \yc$, for red-shifts $z = 0 - 1.75$, in the upper right panel
of Fig. \ref{fig3} shows three separate phases: (i) $\px > 0$ for $y \gtrsim 0.18$, whence the DE is
decelerating ($\wx > 0$), (ii) $-0.7 \lesssim \px < 0$ for $0 < y \lesssim 0.18$, whence the DE is
accelerating but with $\wx > -1$, (iii) $\px \lesssim -0.7$ for $y < 0$, whence we have a
super-accelerating (i.e., phantom-like) DE with $\wx < -1$. The parametric plots of the reconstructed
normalized k-essence Lagrangian components $P_{n_I} = P_{_I}/P_{_I}^{(0)}$ (where $P_{_I}^{(0)} =
P_{_I}\vert_{z=0}, ~I = 0, 1, 2$) against the corresponding normalized temporal field gradients $y_{n_I}
= \yI/\yI^{(0)}$ (where $\yI^{(0)} =\yI\vert_{z=0},~ I = 0, 1, 2$), for $0 \leq z \leq 1.75$, are shown
in the lower panels of Fig. \ref{fig2}. In the reconstructions, we have set for simplicity $\l_{_1}$
and $\l{_2}$ appearing in Eqs. (\ref{rho1-P1}) and (\ref{rho2-P2}) to be zero, whence the cosmological
constant term in Eq. (\ref{ded-pr}) is given by $\l = \l_{_0}$. The pressure $P_{n_1}$ is found to fall
off with increasing values of $y_{n_1}$, whereas the pressures $P_{n_0}$ and $P_{n_2}$ are found to
increase, but with opposite rates, with $y_{n_0}$ and $y_{n_2}$ respectively. Since the best fit values
of both the parameters $\Ab, \Ac$ are positive, the only parameter with a negative value is $\Aa = - 2
\Ab/\b = - 8 \Ab$ (for $\b = 0.25$). Therefore, the fields which contribute to the negative part of $\wx$
are $\f_1$ (due to the negative constant value of EoS parameter $w_{_1}$, but positive definite energy
density $\r_{_1}$) and $\f_0$ (due to the negative parameter $\Aa$, or $\aa$, which drives $\ya = {\dot
\f}_{_0}$ and/or $P_{_0}$ to be negative, in some regime). However, as explained above, the field
$\f_{_0}$ alone is not responsible for the PDL crossing, and it is the field $\f_{_1}$ with Lagrangian
component $P_{_1}$ which effectively causes the super-acceleration ($\wx < -1$).

\section{Conclusions  \la{sec:concl}}

While a PDL crossing DE model involving a single k-essence scalar field suffers from cosmological instabilities
\cite{vik}, a set of multiple k-essence fields are shown to give rise to such PDL crossing, satisfying the conditions
for stability even in the simplistic scenario where the k-fields are mutually non-interacting and the individual
k-essence Lagrangian densities are purely functions of the kinetic terms. Although there are severe restrictions
imposed on the k-field configuration by virtue of the stability criterion, i.e., $\cx^2 \geq 0$, and the
non-super-luminescence ($\cx \leq 1$) of the DE density perturbations, we still find certain field configurations
for which the DE EoS parameter can cross the cosmological constant barrier. We have worked out the general equation
of constraint on the field configuration, with an appropriate choice of the propagation speed, viz., $\cx^2 =$
a constant, lying between $0$ and $1$. While a configuration of two kinetic k-fields, satisfying the constraint
equation, leads to complicated solutions, we have shown that a three k-field configuration gives rise to a
cosmologically stable PDL crossing DE model. For certain choices of some of the parameters, the model is found
to fit well with the latest SNIa Union data compiled by the Supernova Cosmology Project team \cite{kowal}. The
value of the red-shift $z = z_c$ at which the DE equation of state parameter $\wx$ (best fit with the data) crosses
the PDL is found to be $\simeq 0.2$, whereas the best fit present matter density parameter is found to be $\O_{0m}
\simeq 0.265$. These are in good agreement with the values of $z_c$ and $\O_{0m}$ found in the likelihood analysis
done using other model-dependent or model-independent ansatze in the literature \cite{gong,param,pdlc}. On the
other hand, the value of $z$ at which the universe transits from a decelerating phase to an accelerating phase is
found to be $\sim 0.65$, which also agrees fairly well with other independent studies \cite{gong,param,pdlc}.

In the model which we have presented here, the value of the square of the speed of propagation of the DE density
perturbations, $\cx^2$, is as low as $1/6$ throughout the entire course of the DE evolution. We have worked out
the expression for $\cx^2$ in a general setting of multiple k-essence fields considering the most general scalar
type of metric perturbations in a Friedmann-Robertson-Walker background. Although the perturbation equations have
been extremely coupled in the curvature and the isocurvature modes, assuming the coupling to be very weak
in the sub-Hubble scales the speed of propagation of the curvature perturbations in the multi-k-field case is
found to be the one that generalizes the well-known expression for $\cx$ for a single k-essence field found in
ref. \cite{gm}.

The present model exhibits some additional features: Firstly, the dark energy density (best fit with the Union data)
almost tracks the matter density in distant past until exceeding the later at $z \simeq 0.4$, thus providing a
possible resolution to the coincidence problem. Secondly, the best fit dark energy density, extrapolated to future
epochs, does not run away to very high values leading to Big Rip-like singularities even in distant future. Thirdly,
the extrapolated best fit of the dark energy equation of state parameter shows a tendency to turn back from a minimum
negative value ($< -1$) towards the cosmological constant value ($= -1$) in distant future, thus indicating a transient
phantom regime.

We have reconstructed the entire k-essence Lagrangian $P(\ya,\yb,\yc)$, as well as the individual Lagrangian
components $\pI (\yI)$ in terms of the corresponding field gradients $\yI = \dfI$, using the values of the model
parameters best fitting the observational data. The reconstructed total Lagrangian, $P$ is found to be a smooth
function of the total field gradient $y = \sum \yI$, i.e., there are no discontinuities or multi-valuedness in the
parametric plot of $P$ versus $y$ for the red-shift range under consideration, $0 \leq z \leq 1.75$. All the individual
Lagrangian components $\pI$ as functions of the corresponding $\yI$ also show the same smooth nature.  The reconstructed
Lagrangian $P$ is therefore substantially better than the double-valued Lagrangian found for a single kinetic k-essence
field in ref. \cite{aas05} using popular parameterizations, such as that due to Alam {\it et al} \cite{alam} or due to
Chevallier-Polarski-Linder \cite{cpl} and with the old SNIa gold data-set \cite{riess1}.

A few problems that may be looked upon in the context of the present model are as follows: (i) Can we find a
systematic prescription in order to build a stable PDL crossing multiple kinetic k-essence model, instead
of choosing the number of k-fields by hand and implementing the value of $\cx^2$ arbitrarily? (ii) Can we extend
the model to sufficiently higher red-shifts so as to match with the CMB and SDSS data? (iii) Is there a way to
ascertain how far the isocurvature modes of the density perturbations affect the model at higher red-shifts?
(iv) Can we ascertain the status of the future Big Rip singularities, if any, in the context of multi-k-essence,
generically, i.e., by not just resorting to particular models? Or, can we generically ascertain whether the phantom
regime would be eternal or transient for multiple k-essence? Research works addressing some of these questions are
already in progress \cite{ss} and we hope to report on them in near future.

Finally, it should be mentioned that although we have not specifically looked to construct tracking/scaling dark
energy models in the present paper, the possible avoidance of the coincidence problem in our multi-k-essence model
is particularly noteworthy. The value of $\cx$ remains constant and less than the speed of light, unlike in the
coincidence resolving single field k-essence models, which are shown to be affected by the super-luminal propagation
($\cx > 1$) of the density perturbations \cite{bonvin}. However, whether our purely kinetic model specifically belongs
to the class of tracking or scaling k-essence dark energy, is an interesting issue which we hope to circumvent in
future. Assisted accelerated solutions are shown to exist generically for the multi-field k-essence models that have
scaling solutions \cite{tsuji}. The nature of such assisted acceleration may also be of some interest for the purely
kinetic multi-k-essence, that we have studied here, and is worth investigating.

\ack{This work is supported by the Natural Sciences and Engineering Research Council of Canada. The authors
acknowledge useful discussions with S. Odintsov, A. Dasgupta and especially with S. Shankaranarayanan. The
authors thank L. Perivolaropoulos, whose numerical code for the likelihood analysis, using Mathematica,
has been modified and used in this paper. The authors are also thankful to the anonymous referee for important
remarks.}


\section*{Appendix: Speed of propagation of multiple k-essence cosmological perturbations}

Let us consider the system of $N$ mutually non-interacting k-essence fields coupled to gravity. The
gravitational field equations for such a system are
\be \la{feq}
G_{\m\n} ~=~ \k^2 T_{\m\n} \quad, \qquad T_{\m\n} ~=~ \sum_{I=0}^{N-1} T_{\m\n}^I \, ,
\ee
where $T_{\m\n}^I , (I = 0, \dots, N-1)$ are the energy-momentum tensor components due to the individual
k-fields ${\mathbf \F} = (\f_{_0}, \cdots, \f_{_{N-1}})$, with kinetic terms ${\mathbf X} = (X_{_0},
\cdots, X_{_{N-1}})$, where $\xI = - \fr 1 2 g^{\m\n} \pa_\m \fI \pa_\n \fI$.

The line element with the most general scalar type of perturbations, in an arbitrary gauge, is given by
\cite{bardeen}:
\bea \la{pert-metric}
ds^2 ~= &&- \le[1 + 2 A(x^i,t)\ri] dt^2 ~+~ 2 a(t) ~\pa_i B(x^i,t)~ dx^i dt \nn\\
&&+~ a^2 (t) \le[\le(1 + 2 \psi(x^i,t)\ri) \d_{ij} ~+~ \nabla_j \pa_i E(x^i,t)\ri] dx^i dx^j \, ,
\eea
where $a(t)$ is the cosmological scale factor, and $A(x^i,t), B(x^i,t), \psi(x^i,t)$ and $E(x^i,t)$ are
the perturbed order variables. The perturbed energy-momentum tensor components are given by
\bea \la{pert-em}
T_0^0 = - \le({\bar \r} + \d\r\ri) \, , \quad T_i^0 = - \fr a k \le(\r + p\ri) \pa_i v \, , \nn\\
T_i^j = \le({\bar p} + \d p\ri) \d_i^j + \fr 1 {k^2} \le(\nabla_i \nabla^j + \fr 1 3 \d_i^j\ri) \p \, ,
\eea
where ${\bar \r}$ and ${\bar p}$ are the total background energy density and isotropic pressure respectively,
$\r = {\bar \r} + \d\r$ and $p = {\bar p} + \d p$ are the perturbed total energy density and isotropic pressure
respectively, $v$ is the velocity (or, flux) related variable, $\p$ is the anisotropic stress, and $k$ is the
co-moving wave-number. Decomposing into the components due to individual k-fields, we have
\bea \la{k-decomp}
\fl && {\bar \r} = \sum_I {\bar \r}_{_I} \quad, \quad {\bar p} = \sum_I {\bar P}_{_I} \quad, \quad \d\r = \sum_I
\d\rI \quad, \quad \d p = \sum_I \d\pI \quad, \nn\\
\fl && \r = \sum_I \rI \quad, \quad p = \sum_I \pI \quad, \quad \le(\r + p\ri) v = \sum_I \le(\rI + \pI\ri) \vI
\quad, \quad \p = \sum_I \piI \, .
\eea

The background field equations are the Friedmann equations and the equations of continuity of the individual
scalar fields:
\bea
&&H^2 ~\equiv~ \le(\fr{\dot a} a\ri)^2 ~=~ \fr{\k^2} 3 ~ \r \, , \la{sf-feq-1}\\
&&{\dot H} ~\equiv~ \fr{\ddot a} a ~-~ \fr{{\dot a}^2}{a^2} ~=~ -~ \fr{\k^2} 2 \le(\r + p\ri) \, , \la{sf-feq-2}\\
&&\drI ~+~ 3 H \le(\rI + \pI\ri) ~=~ 0 \quad, \qquad \mbox{for all}~~ I = 0, \dots, N-1 \, , \la{sf-feq-3}
\eea

The equations for the cosmological perturbations
\be \la{cp-eq}
\d G_\a^\b ~=~ \k^2 \d T_\a^\b  \, ,
\ee
can be decomposed as
\bea
\fl H \xi ~-~ \fr{k^2}{a^2} \psi ~=~ - \fr{\k^2} 2~ \d\r \qquad &:& \d G_0^0 ~~\mbox{component} \, ,
\quad  \la{cp-eqs-1}\\
\fl \xi ~-~ \fr{k^2}{a^2} \chi ~=~ \fr{3 \k^2} 2 ~ \fr a k \le(\r + p\ri) v \qquad &:& \d G_i^0
~~\mbox{component} \, , \quad  \la{cp-eqs-2}\\
\fl {\dot \chi} ~+~ H \chi ~=~ A ~+~ \psi \qquad &:& \d G_i^j - \fr 1 3 \d_i^j \d G_k^k
~~\mbox{component} \, , \quad  \la{cp-eqs-3}\\
\fl {\dot \xi} ~+~ 2 H \xi ~+ \le(3 {\dot H} - \fr{k^2}{a^2}\ri) A ~=~ \fr{\k^2} 2 \le(\d\r + 3 \d p\ri)
\qquad &:& \d G_k^k - \d G_0^0 ~~\mbox{component} \, , \quad  \la{cp-eqs-4}
\eea
where
\be \la{chi-xi}
\chi ~=~ a \le(B + a {\dot E}\ri) \quad, \qquad \xi ~=~ 3 \le(H A - {\dot \psi}\ri) +~ \fr{k^2}{a^2}~ \chi \, ,
\ee
and note that $\p = 0$ for a scalar field system in RW background. The quantities $\psi, \chi$ and $\xi$ correspond
respectively to the three-space curvature, the shear, and the perturbed expansion of the normal-frame vector field
\cite{hwang,hwangnoh1}. Eq. (\ref{cp-eqs-1}) gives the Arnowitt-Deser-Misner (ADM) energy constraint, Eq. (\ref{cp-eqs-2})
gives the momentum constraint, Eq. (\ref{cp-eqs-3}) is the trace-free part of the ADM propagation, and Eq. (\ref{cp-eqs-4})
is the Raychaudhuri equation \cite{hwangnoh1}. In addition to the above equations, we have the energy and momentum
conservation equations given by
\bea
\d \drI ~+~ 3 H \le(\d\rI + \d\pI\ri) &=& \le(\xi - 3 H A - \fr k a \vI\ri) \le(\rI + \pI\ri) \quad, \la{cp-consv-1}\\
\fr{\pa_t \le[a^4 \le(\rI + \pI\ri) \vI\ri]}{a^4 \le(\rI + \pI\ri)} &=& \fr k a \le(A + \fr{\d\pI}{\rI +
\pI}\ri) \, . \la{cp-consv-2}
\eea

The entropic perturbations are given by
\be \la{ep}
\eI ~=~ \d \pI ~-~ \cI^2 \d \rI \quad, \qquad e ~=~ \d p ~-~ c_s^2 \d \r \, ,
\ee
where $\cI^2 = \dpI/\drI$ and $c_s^2 = {\dot p}/{\dot \r}$.

Under a coordinate transformation
\be \la{coord-tr}
\fl \qquad x^\a \to x^\a + q^\a \quad, \qquad q^\a = \le(q^0, q^i\ri) \quad; \quad q^0 = \d t \quad, \quad q^i = \d^{ij}
\pa_j \d x \, ,
\ee
the metric perturbations transform as
\be \la{mp-tr}
\fl \quad A \to A - \d {\dot t} \quad, \quad B \to B - \fr{\d t} a + a \d {\dot x} \quad, \quad \psi \to \psi - H \d t
\quad, \quad E \to E - \d x \, ,
\ee
whereas the field perturbations transform as
\be \la{fp-tr}
\d \fI \to \d \fI - \dfI \d t \, .
\ee
Since we are always free to choose $\d t$ and $\d x$, we can impose two gauge conditions on the functions $A, B,
\psi, E$ or $\d\r$ or $\pa_i v$. We choose to work in the commonly used longitudinal gauge in which $B = E = 0$.
Then from Eq. (\ref{chi-xi}) $\chi = 0$, whence Eq. (\ref{cp-eqs-3}) gives $A = - \psi$, and the expression for
$\xi$ in Eq. (\ref{chi-xi}) reduces to
\be \la{xi-def}
\xi ~=~ -~ 3 \le({\dot \psi} + H \psi\ri) \, .
\ee

\bigskip
\noindent
{\large \bf \underline{Curvature Perturbations}:}
\bigskip

The so-called gauge-invariant co-moving curvature perturbation variable is given by \cite{luk}
\be \la{cp-var}
\F ~=~ \psi ~-~ \fr a k H v \, .
\ee
From Eqs. (\ref{cp-eqs-1}), (\ref{cp-eqs-2}), (\ref{ep}) and (\ref{xi-def}), one derives the equations for
the curvature perturbations as
\bea
&&{\dot \psi} ~+~ H \psi ~=~ - \fr{\k^2} 2 ~ \fr a k \le(\r + p\ri) v \quad, \la{eqs-cp-1}\\
&&\fr{H^2}{\le(\r + p\ri) a^3} \pa_t \le(\fr{a^3} H ~\d \r_v\ri) ~=~ \fr{k^2}{a^2} \F \, , \la{eqs-cp-2}
\eea
where the gauge invariant variable $\d \r_v$ is defined by
\be \la{delta-rhov}
\d \r_v ~=~ \d \r - \fr a k {\dot \r} v ~=~ - \fr 1 {c_s^2} \le(\fr{\r + p} H {\dot \F} + e\ri) \, .
\ee

\bigskip
\noindent
{\large \bf \underline{Isocurvature Perturbations}:}
\bigskip

The total entropic perturbation can be decomposed as \cite{hwangnoh1,kodamasasaki}
\be \la{ep-comp}
\fl e = e_r + \sum_I \eI \quad, \quad e_r = \sum_I \le(\cI^2 - c_s^2\ri) \d \rI =
\fr 1 2 \sum_{I,J} \fr{\le(\rI + \pI\ri) \le(\rJ + \pJ\ri)}{\r + p} \cIJ^2 \SIJ \, ,
\ee
where $\cIJ^2 = \cI^2 - \cJ^2$, and the so-called gauge-invariant isocurvature perturbation variable
$\SIJ$ is defined as
\be \la{ip-var}
\SIJ ~=~ \fr{\d \rI}{\rI + \pI} ~-~ \fr{\d \rJ}{\rJ + \pJ} \, .
\ee
From Eqs. (\ref{cp-consv-1}), (\ref{cp-consv-2}) and (\ref{ep}) one obtains the equations for the
isocurvature perturbations
\bea
\fl {\dot S}_{_{IJ}} &=& - \fr k a \vIJ - 3 H \eIJ \quad, \qquad \vIJ = \vI - \vJ \quad,
\quad \eIJ = \fr{\eI}{\rI + \pI} - \fr{\eJ}{\rJ + \pJ} \quad, \la{eqs-ip-1}\\
\fl {\ddot S}_{_{IJ}} &+& 2 H {\dot S}_{_{IJ}} ~-~ \fr{3H} 2 \le[\cbIJ^2 {\dot S}_{_{IJ}} ~+~
\sum_K \fr{\r_{_K} + P_{_K}}{\r + p} \le({\dot S}_{_{IK}} + {\dot S}_{_{JK}}\ri)\ri] \nn \\
\fl &+& \fr{k^2}{2 a^2} \le[\cbIJ^2 \SIJ ~+~ \cIJ^2 \sum_K \fr{\r_{_K} + P_{_K}}{\r + p} \le(S_{_{IK}}
+ S_{_{JK}}\ri)\ri] \nn \\
\fl &=& -~ 3 H {\dot e}_{_{IJ}} ~-~ 3 \le({\dot H} + 2 H^2\ri) \eIJ ~+~ \fr {9H^2} 2 \le[\cbIJ^2 \eIJ
~+~ \cIJ^2 \sum_K \fr{\r_{_K} + P_{_K}}{\r + p} \le(e_{_{IK}} + e_{_{JK}}\ri)\ri] \nn\\
\fl &-& \fr{k^2}{a^2} \le(\cIJ^2 \fr{\d \r_v}{\r + p} ~+~ \eIJ\ri) \, , \la{eqs-ip-2}
\eea
where $\cbIJ^2 = \cI^2 + \cJ^2$.

Eqs. (\ref{eqs-cp-1}), (\ref{eqs-cp-2}) and (\ref{eqs-ip-1}), (\ref{eqs-ip-2})  form an extremely
coupled set of differential equations in the curvature and isocurvature modes, very difficult to solve
in general for any system of multiple fields. However, leaving aside the isocurvature terms, one
may still work out the effective speed of propagation of curvature perturbations in such a system. In
fact, in the large scale limit ($k/a \to 0$), the curvature mode couples very weakly to the isocurvature
ones, as is evident from the term $(k^2/a^2) \cIJ \d \r_v/(\r + p)$ on the right hand side
of Eq. (\ref{eqs-ip-2}). For a detailed discussion on the issue of decoupling of isocurvature
and curvature modes in multiple fluid or field systems, see ref. \cite{hwangnoh1}.

Now, in the case of mutually non-interacting multiple k-essence scalar fields, we have the
expressions for the total energy density and pressure
\bea
\r &\equiv& \rx = \sum_{I=0}^{N-1} \rI (\fI,\xI) = \sum_{I=0}^{N-1} 2 \le[\xI \fr{\pa \pI}
{\pa \xI} - \pI (\fI,\xI)\ri] \, , \la{den-pr-1} \\
p &\equiv& \px = \sum_{I=0}^{N-1} \pI (\fI,\xI) \, . \la{den-pr-2}
\eea
Using the continuity equations for the fields ${\mathbf \F} = (\f_{_0}, \cdots, \f_{_{N-1}})$, and
the fact that $\xI = \dfI^2/2$ in RW background, we obtain the expressions for the perturbed variables
$\d \r$ and $\d p$ as
\bea
\d \r &=& \sum_{I=0}^{N-1} \d \rI \quad , \qquad \d \rI ~=~ \fr{\drI}{3 H \ctI^2} \le(3 H
\ctI^2 \sigI ~-~ {\dot \s}_{_I} ~-~ \psi\ri) \quad, \la{den-pr-pert-1}\\
\d p &=& \sum_{I=0}^{N-1} \d \pI \quad , \qquad \d \pI ~=~ \fr{\drI}{3 H} \le(3 H \cI^2 \sigI
~-~ {\dot \s}_{_I} ~-~ \psi\ri) \quad, \la{den-pr-pert-2}
\eea
where we have defined $\sigI = \d \fI/\dfI$ and $\ctI^2 = \fr{\pa \pI}{\pa \xI}/\fr{\pa \rI}
{\pa \xI}$.

The expressions for $\vI$ and $v$, obtained from the $0-i$ components of the energy-momentum
tensors, are given by
\be \la{vi}
\vI ~=~ \fr k a ~\sigI \quad, \qquad v ~=~ \fr k {a {\dot \r}} \sum_{I=0}^{N-1} \drI \sigI \, .
\ee

Defining further $\s = \sum \sigI$ and $\sigIJ = \sigI - \sigJ$, we get from Eqs. (\ref{cp-var}),
(\ref{eqs-cp-1}) and (\ref{delta-rhov})
\bea
&& \F ~=~ \psi ~-~ \fr{H \s} N ~+~ {\mathcal O} (\sigIJ) \quad, \la{cp-var-eqs-1}\\
&& {\dot \psi} ~+~ H \psi ~=~ \fr{{\dot H} \s} N ~+~ {\mathcal O} (\sigIJ) \quad, \la{cp-var-eqs-2}\\
&& \d \r_v ~=~ \fr{\dot F}{3 H^2} \sum_{I=0}^{N-1} \fr{\drI}{\ctI^2} ~+~ {\mathcal O} (\sigIJ)
\quad, \la{cp-var-eqs-3}
\eea
where $N$ is the total number of fields and for the reasons mentioned earlier we have left aside
the terms depending only on $\sigIJ$ (and/or its time derivatives), which give rise to the isocurvature
modes.

The equation (\ref{eqs-cp-1}) for the curvature mode can now be expressed as
\be \la{curv-eq}
\fr{H^2}{\le(\rx + \px\ri) a^3} \pa_t \le[\fr{\le(\rx + \px\ri) a^3}{\cx^2 H^2} ~{\dot \F}\ri] +~
\fr{k^2}{a^2} \F ~=~ {\mathcal O} (\sigIJ) \quad,
\ee
where
\be \la{ceff-sq}
\cx^2 ~=~ {\dot \r} \le[\sum_{I=0}^{N-1} \fr{\drI}{\ctI^2}\ri]^{-1} ~=~ (\rx + \px) \le[ \sum_{I=0}^{N-1}
\fr{\rI + \pI}{\ctI^2} \ri]^{-1} \, ,
\ee
is the effective speed of propagation of the curvature mode of scalar perturbations \cite{ss}. For a
single k-field ($N=1, I=0$), Eq. (\ref{ceff-sq}) reduces to the well-known result $\cx^2 = \tilde{c}_{_0}^2
= \pa P_{_0}/\pa X_{_0}$ \cite{gm}, whereas for purely kinetic multiple k-essence fields, we have $\ctI^2 =
\cI^2 = \dpI/\drI$, whence $\cx^2 = {\dot \r}/\sum (\drI^2/\dpI)$. If any of the fields is non-propagating,
i.e., $\cI^2 = 0$ for some value of $I$, then $\cx^2 = 0$.

Under the transformation
\be
u ~=~ \zeta \F \quad, \qquad \zeta ~=~ \fr{a \sqrt{\rx + \px}}{\cx H} \, .
\ee
the above equation (\ref{curv-eq}) reduces to the familiar form \cite{gm,hwang,bardeen,luk}
\be
u'' ~+~ \le(\cx^2 k^2 ~-~ \fr{\zeta'}{\zeta}\ri) u ~=~ {\mathcal O} (\sigIJ) \quad,
\ee
where the prime $\{'\}$ here denotes differentiation with respect to the conformal time $\eta$ ($= \int
dt/a(t)$).

\end{document}